\documentclass[iop,apj]{emulateapj}

\usepackage{graphicx,natbib,amsmath,amssymb}
\bibpunct[, ]{(}{)}{;}{a}{}{,}
\newcommand{\simgt}{\hbox{\rlap{\raise 0.425ex\hbox{$>$}}\lower 0.65ex\hbox{$\sim$}}}
\newcommand{\simlt}{\hbox{\rlap{\raise 0.425ex\hbox{$<$}}\lower 0.65ex\hbox{$\sim$}}}

\newcommand{\msun}{M$_\odot$}
\newcommand{\lsun}{L$_\odot$}
\newcommand{\mdust}{$M_{\rm dust}$}
\newcommand{\mstar}{$M_\star$}

\newcommand{\msunyr}{M$_\odot$ yr$^{-1}$}

\submitted{Received 2013 November 25; accepted 2014 January 21}
\journalinfo{The Astrophysical Journal Letters, in press (2014)}

\shorttitle{Relation between dust and star formation}
\shortauthors{Hjorth et al.}

\begin{document}

\title{
Shaping the dust mass -- star-formation rate relation
}


\author{ 
Jens Hjorth\altaffilmark{1},
Christa Gall\altaffilmark{1,2,3},
and
Micha{\l} J. Micha{\l}owski\altaffilmark{4,5,6}
}

\altaffiltext{1}{Dark Cosmology Centre, Niels Bohr Institute, 
University of Copenhagen, Juliane Maries Vej 30, DK-2100 Copenhagen \O, 
Denmark} 
\altaffiltext{2}{Department of Physics and Astronomy, 
Aarhus University, Ny Munkegade 120, DK-8000 Aarhus C, Denmark}
\altaffiltext{3}{NASA, Goddard Space Flight Center, 8800 Greenbelt Road,
Greenbelt, MD 20771}
\altaffiltext{4}{Sterrenkundig Observatorium, Universiteit Gent, 
Krijgslaan 281-S9, 9000, Gent, Belgium}
\altaffiltext{5}{SUPA (Scottish Universities Physics Alliance), Institute for 
Astronomy, University of Edinburgh, Royal Observatory, Edinburgh, EH9 3HJ, UK}
\altaffiltext{6}{FWO Pegasus Marie Curie Fellow}


\begin{abstract}
There is a remarkably tight relation between the observationally inferred dust 
masses and star-formation rates (SFRs) of SDSS galaxies, \mdust $\propto$ 
SFR$^{1.11}$ \citep{2010MNRAS.403.1894D}. Here we extend the \mdust--SFR 
relation to the high end and show that it bends over at very large SFRs 
(i.e., dust masses are lower than predicted for a given SFR). We identify 
several distinct evolutionary processes in the diagram:
(1) A {\em star-bursting phase} in which dust builds up rapidly at early times.
The maximum attainable dust mass in this process is the cause of the bend-over 
of the relation. A high dust-formation efficiency, a bottom-light initial mass 
function, and negligible supernova shock dust destruction are required to 
produce sufficiently high dust masses.
(2) A {\em quiescent star-forming phase} in which the subsequent parallel 
decline in dust mass and SFR gives rise to the \mdust--SFR relation, through 
astration and dust destruction. The dust-to-gas ratio is approximately 
constant along the relation. We show that the power-law slope of the 
\mdust--SFR relation is inversely proportional to the global 
Schmidt--Kennicutt law exponent (i.e., $\sim 0.9$) in simple chemical 
evolution models.
(3) A {\em quenching phase} which causes star formation to drop while the 
dust mass stays roughly constant or drops proportionally. Combined with
{\em merging}, these processes, as well as the range in total baryonic mass,
give rise to a complex population of the diagram which adds significant
scatter to the original \mdust--SFR relation.
(4) At very high redshifts, a population of galaxies located
significantly below the local relation is predicted.
\end{abstract}





\section{Introduction\label{introduction}}

Galactic scaling relations such as the `main sequence of star formation' 
\citep[the relation between star-formation rate and stellar mass, 
see][]{2007ApJ...660L..43N,2007ApJ...670..156D,2011ApJ...742...96W} provide 
important constraints on galaxy evolution models and the physical processes 
involved.

\begin{figure}
\plotone{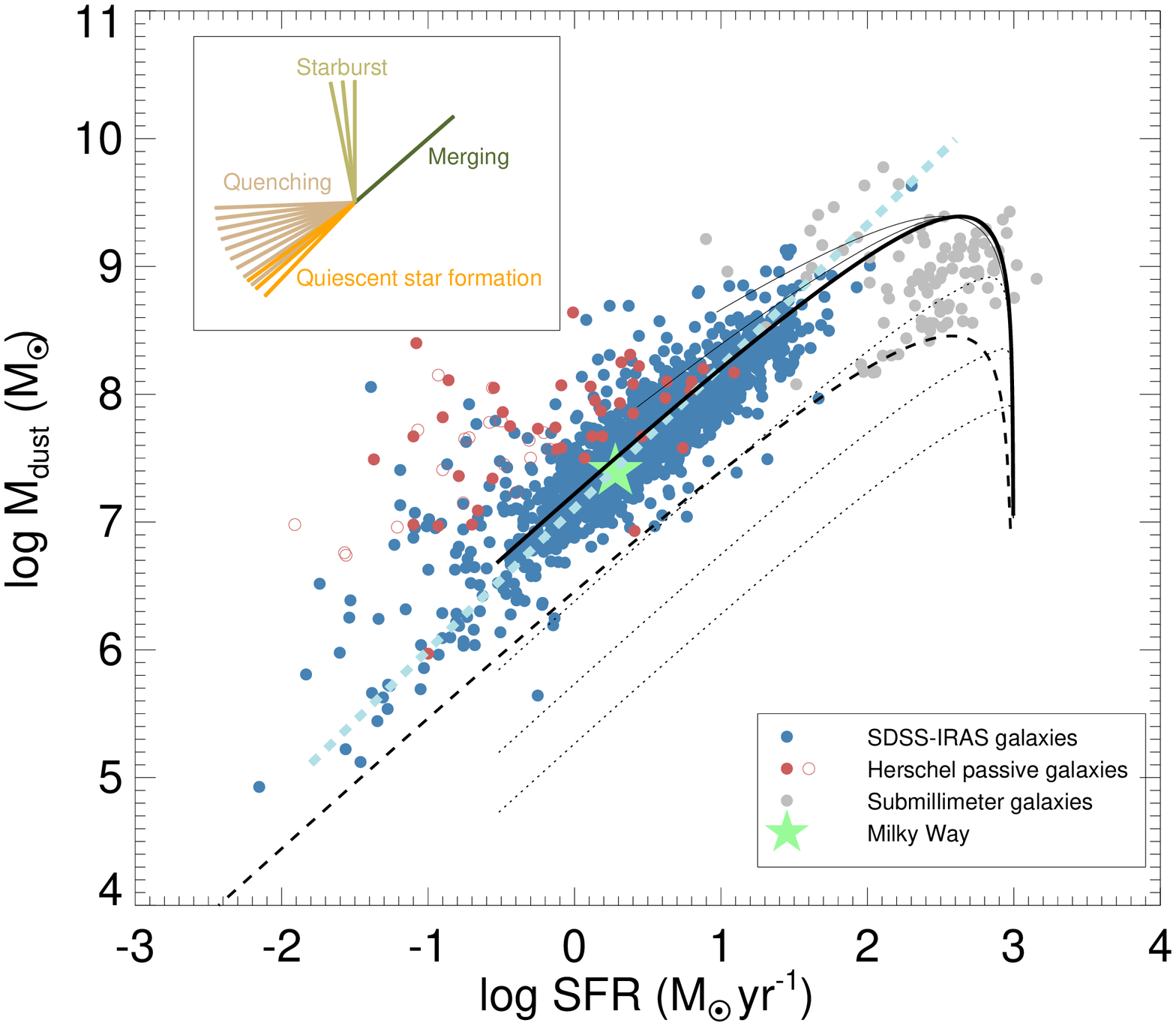}
\caption{The dust mass -- star-formation rate relation.
Blue symbols: SDSS--IRAS galaxies from \citet{2010MNRAS.403.1894D}.
Green star: Milky Way \citep{Hjorth14}.
Red symbols: ETGs (filled), passive spirals (open) from 
\citet{2012MNRAS.419.2545R}.
Grey symbols: SMGs from
\citet{2013arXiv1310.6362S}.
The thick dashed pale-blue line is the \citet{2010MNRAS.403.1894D} relation.
All SFRs have been computed assuming a Chabrier IMF.
Overplotted are models with initial SFRs equivalent to
1000 \msunyr\ for a Chabrier IMF, and gas mass of 
$3 \times 10^{11}$ \msun, as described in 
Section~3. 
The thick solid curve is for a bottom-light IMF, no SN dust destruction, and 
$k=0.9$. 
Thin solid curves are for $k=1$ and $k=1.2$.
The thick dashed curve is for a Salpeter IMF. 
The dotted curves are for $M_{\rm cl}$ of 100, 500, 1500 \msun.
The different physical processes shaping the diagram are shown as directions 
in the upper left corner.
\label{f:dacunha}
}
\end{figure}

One such relation reflects that massive star-forming galaxies contain large 
amounts of dust while old elliptical (red and dead) galaxies do not. This 
qualitative notion was quantified by \citet{2010MNRAS.403.1894D} who showed 
that for SDSS galaxies there is a tight \mdust--SFR relation, \mdust\ $\sim$ 
SFR$^{1.11\pm0.01}$. The \mdust--SFR relation presented by 
\citet{2010MNRAS.403.1894D} consists of 3258 low-redshift SDSS galaxies with 
complementary data from GALEX, 2MASS and IRAS. The 1653 data points with the 
highest signal-to-noise ratio are reproduced in Figure~\ref{f:dacunha} 
(blue data points).

Naively, one might interpret this to be a causal relation, either due to star 
formation induced by the prior presence of dust as seeds for star formation 
(although this poses the question of where the initial amounts of dust come 
from in the first place) or as a direct consequence of dust produced following 
massive star formation (or, rather, death). However, \cite{2010MNRAS.403.1894D} 
advocated a different scenario in which the relation emerges as a consequence 
of the parallel decline of dust and SFR due to the decreasing gas mass 
available for star formation as time goes on and a simultaneous decline in 
total dust mass due to dust destruction.

To shed further light on the \mdust--SFR relation we here populate the 
\cite{2010MNRAS.403.1894D} observational diagram with additional classes of
galaxies. In particular, in 
Section~2 
we extend it to higher SFR and \mdust\ 
by including sub-millimeter galaxies (SMGs). 
These galaxies bring important new insight into the physical origin of the 
relation. In
Section~3
we model the extended relation using simple chemical evolution models for 
massive galaxies, both from a simple analytical perspective and using full 
numerical models. In particular, we show that the slope of the relation can 
be related to the exponent of the Schmidt--Kennicutt star-formation law, 
relating the SFR to the available gas reservoir. Finally, in
Section~4, 
we discuss how starbursts, quiescent star formation and quenching of the star 
formation cause galaxies to move around in the diagram and contribute to the 
significant scatter revealed here.

\section{The \mdust--SFR relation\label{relation}}

We supplement the original \citet{2010MNRAS.403.1894D} \mdust--SFR relation 
with the {\em Herschel} early-type galaxies (ETGs) and passive spirals 
studied by \citet{2012MNRAS.419.2545R}. These are plotted in 
Figure~\ref{f:dacunha} as red filled (ETGs) and open (spirals) circles. 
The Milky Way data point is taken from \citet{Hjorth14}.
For SMGs we use recent ALMA based data points from
\citet{2013arXiv1310.6362S}.

The \citet{2010MNRAS.403.1894D} and \citet{2012MNRAS.419.2545R} data points are 
for a \citet{2003PASP..115..763C} initial mass function (IMF) and a dust mass 
absorption coefficient of $0.77$ cm$^2$ g$^{-1}$ at 850 $\mu m$, as implemented 
in MAGPHYS \citep{2008MNRAS.388.1595D}. The \citet{2013arXiv1310.6362S} SFR was 
obtained by multiplying the far-infrared luminosity by 
$10^{-10}$ \msun\ \lsun$^{-1}$ yr$^{-1}$ (appropriate for a Chabrier IMF; 
the UV contribution to the total SFR in SMGs is negligible). The quoted dust 
masses were multiplied by 1.5/0.77 to rescale to the same dust mass absorption 
coefficient. We note that while the literature data have been scaled to the 
same IMF and dust mass absorption coefficient, systematic offsets in the 
different subsamples may remain, due to different wavelength coverage, 
redshift, and methods of analysis used. For example, the lack of 
long-wavelength ($>100$ $\mu$m) data in \citet{2010MNRAS.403.1894D} may bias 
dust temperatures high and hence dust masses low \citep{2012MNRAS.427..703S}.

The \citet{2010MNRAS.403.1894D} relation (shown as a thick pale-blue dashed 
line) does not extend to very high SFRs and dust masses. Extrapolating the 
relation to high SFRs shows that observed dust masses are about an order of 
magnitude smaller than predicted (this effect would be even stronger if the 
\citet{2010MNRAS.403.1894D} dust masses are underestimated). It is also evident 
that the relation is not as sharply defined as suggested by the SDSS galaxies 
alone. We discuss these effects below.

\section{Evolution of dust and star formation in massive 
galaxies\label{chemical}}

Chemical evolution models can be used to study the temporal evolution of dust, 
gas, abundance distribution of elements, stellar mass and metallicity in a 
galaxy. The models are governed by processes regulating the evolution of a 
galaxy and, thus, are useful to study not only the chemical history but also 
to trace the SFR and the effects of the IMF, gas flows, and dust destruction 
and growth processes. Chemical evolution models have been applied to diverse 
types of galaxies, such as the Milky Way \citep[see][and references 
therein]{1998ApJ...501..643D, 2008A&A...479..669C, 2008A&A...479..453Z},
high-redshift galaxies 
\citep[e.g.,][]{2001MNRAS.328..223E,2003MNRAS.343..427M} or high-$z$ quasars 
\citep[e.g.,][]{2007ApJ...662..927D, 
2011A&A...528A..14G}.
 
\subsection{Basic equations\label{basic}}

In this Letter we model a galaxy as a homogeneous entity (no spatial dependence 
of its properties) and with no infalling or outflowing material (closed box 
approximation). Efficient supernova (SN) dust production is assumed, while dust 
grain growth in the interstellar medium (ISM) is not explicitly considered. 
Relevant rates are assumed to be independent of metallicity. The model is 
described in detail in \citet{2011A&A...528A..13G}, whose notation we adopt.

The rate of change of the total dust mass is 
\begin{equation}
\frac{dM_{\mathrm{d}}}{dt} = 
E_{\mathrm{d,SN}}(t)+E_{\mathrm{d,AGB}}(t)-E_{\mathrm{D}}(t),
\end{equation}
where $E_{\mathrm{d,SN}}(t)$ and $E_{\mathrm{d,AGB}}(t)$ are the dust injection 
rates of SNe and asymptotic giant branch (AGB) stars, respectively. 
The SN dust production rate is
\begin{equation}
E_{\rm d,SN}(t) 
=\int_{m_{\rm L}}^{m_{\rm U}} Y_{\rm Z}(m) \, \epsilon_{\rm SN}(m) \, 
\psi(t-\tau(m)) 
\phi(m)dm,
\end{equation}
where $m_{\rm L}$ and $m_{\rm U}$ are the lower and upper mass limits for
stars exploding as SNe,
$Y_{\rm Z}(m)$ is the mass of ejected metal yields per SN,
$\epsilon_{\rm SN}(m)$ is the SN dust production efficiency as defined in 
\citet{2011A&ARv..19...43G},
$\psi(t)$ is the SFR, 
$\tau(m)$ is the main sequence lifetime of a star with mass $m$, and
$\phi(m)$ is the IMF, normalized in the interval 
$[m_{1}, m_{2}]$ as $\int_{m_{1}}^{m_{2}}  m \,  \phi(m) \, \mathrm{d}m = 1$,  
with $m_1$ and $m_2$ being the low and high mass cutoffs for the adopted IMF, 
respectively. The rate of dust destruction due to astration and shocks in 
the interstellar medium 
\cite[e.g.,][]{1989IAUS..135..431M,1996ApJ...469..740J}, respectively, is 
defined as
\begin{equation}
E_{\mathrm{D}}(t) = \eta_d(t)(\psi(t)+M_{\mathrm{cl}}R_{\mathrm{SN}}(t)),
\end{equation}
where $M_{\mathrm{cl}}$ is the mass of interstellar material swept up and 
cleared of dust by a single SN, and
\begin{equation}
R_{\mathrm{SN}}(t) 
=\int_{m_{\mathrm{L}}}^{m_{\mathrm{U}}} \psi(t-\tau) \phi(m) \mathrm{d}m
\end{equation}
is the SN rate. The dust-to-ISM ratio is
\begin{equation}
\eta_{\mathrm{d}}(t) = \frac{M_{\mathrm{d}}(t)}{M_{\mathrm{ISM}}(t)},
\end{equation}
where
\begin{equation}
M_{\mathrm{ISM}}(t)=M_{\mathrm{d}}(t)+M_{\mathrm{g}}(t).
\end{equation}

The equation for the evolution of the gas mass is
\begin{equation}
\frac{dM_{\mathrm{g}}}{dt} = E_{\mathrm{g}}(t)+ \eta_{\mathrm{d}}(t)M_{\mathrm{cl}}R_{\mathrm{SN}}(t)- 
\psi(t) \, (1-\eta_{\mathrm{d}}(t)),
\end{equation}
where $E_{\mathrm{g}}(t)$ is the rate of gaseous material returned to the ISM.

Finally, we assume a global star-formation law, inspired by the  
Schmidt--Kennicutt law \citep{1959ApJ...129..243S,1998ApJ...498..541K},
\begin{equation}
\psi(t) = \psi_{\rm ini} \left ( \frac{M_{\mathrm{ISM}}(t)}{M_{\rm ini}}\right ) ^k,
\label{schmidt}
\end{equation}
\citep[Equation (2) in][]{2011A&A...528A..13G}, 
where $\psi_{\rm ini}$ and $M_{\mathrm{ini}}$ are the initial SFR and gas mass,
and $k$ is the global Schmidt--Kennicutt exponent, usually taken to be 1.0 or 
1.5 \citep{2007ApJ...662..927D,2008A&A...479..669C}.

\subsection{Approximations and simplifications\label{simplifications}}

We next introduce several simplifications which will allow us to capture the 
essentials (but not the details) of the chemical evolution equations:

We assume that SN yields are released into the ISM instantly 
after the progenitor star is born, i.e., $\tau\approx 0$, in which case
\begin{equation}
R_{\mathrm{SN}}(t) =\gamma \psi(t),
\end{equation}
where
\begin{equation}
\gamma = \int_{m_{\mathrm{L}}}^{m_{\mathrm{U}}} \phi(m) \mathrm{d}m
\end{equation}
is the SN rate to SFR ratio.

We ignore the contribution from AGB stars 
\citep[e.g.,][]{2007ApJ...662..927D,2010A&A...522A..15M,2011A&ARv..19...43G}
or other sources to the evolution of the dust mass, so
\begin{equation}
\frac{dM_{\mathrm{d}}}{dt} = E_{\mathrm{d,SN}}(t)-E_{\mathrm{D}}(t),
\end{equation}
with
\begin{equation}
E_{\rm d,SN}(t) = \mu_D \psi(t)
\end{equation}
and
\begin{equation}
E_{\mathrm{D}}(t) = \beta \eta_d(t)\psi(t),
\end{equation}
where $\mu_D$ is the dust productivity \citep{2011A&ARv..19...43G}, and
\begin{equation}
\beta = 1+\gamma M_{\mathrm{cl}}.
\end{equation}
The resulting equation for the evolution of the dust mass is
\begin{equation}
\frac{dM_{\mathrm{d}}}{dt} = (\mu_D - \beta \eta_d(t)) \psi(t).
\end{equation}

We also assume that $\eta_d \ll 1$ 
\citep[e.g.,][]{2005pcim.book.....T,2011A&A...528A..13G}
and that $E_{\mathrm{g}}(t) \propto \psi(t)$, so the evolution of gas mass is
\begin{equation}
\frac{dM_{\mathrm{g}}}{dt} = - \delta \psi(t),
\end{equation}
where $\delta$ is a constant of proportionality close to unity, and
\begin{equation}
\psi(t)=\alpha M_g(t)^k; \ \ \ \ \ \ \ 
\alpha=\frac{\psi_{\rm ini}}{M_{\rm ini}^k} .
\label{late1}
\end{equation}

We study IMFs with $\phi(m)\propto m^{-2.35}\exp(-m_{ch}/m)$ with 
$m_{\rm L}=8$ \msun\ and $m_{\rm U}=40$ \msun,
$m_{\rm 1}=0.1$ \msun\ and $m_{\rm 2}=100$ \msun,
and maximum SN dust production efficiency \citep{2011A&ARv..19...43G},
corresponding to an average dust yield per SN of $\sim 0.3$--0.4 \msun\
\citep{Hjorth14}. In particular, we study Salpeter ($m_{ch}=0$) and 
bottom-light ($m_{ch}=10$ \msun) IMFs for which the conversions to Chabrier 
IMF SFRs are 1/1.8 and 3.0 \cite[see, for example,][]{2011ApJ...738...36D}. 
For a bottom-light (Salpeter) IMF, the approximate values of the parameters 
entering are 
$\gamma\approx 0.02$ (0.007) \msun$^{-1}$,
$\delta\approx0.83$,
$\mu_D\approx0.018$ (0.003),
$M_{\rm cl}=0$--1500 \msun, and
$\beta=1$--30.

\subsection{Analytical results\label{analytical}}

We next investigate simple analytical limiting cases at early and late times 
resulting from the above set of equations.

\subsubsection{Early times: The starburst limit\label{early}}

At early times, not much gas has been consumed, so $M_g\approx M_{\rm ini}$ and
$\psi\approx \psi_{\rm ini}$. Thus
\begin{equation}
\frac{dM_{ \rm d}}{dt} = \mu_D \psi_{\rm \rm ini} - 
\frac{ \beta \psi_{\rm ini} } {M_{\rm ini}} M_{ \rm d}(t)
\end{equation}
and so
\begin{equation}
M_{ \rm d}(t) = \frac{\mu_D}{\beta} M_{\rm ini}
\left (1-\exp\left ( {-\beta \frac{\psi_{\rm ini}}{M_{\rm ini}} t} 
\right ) \right ).
\end{equation}
In other words the maximum attainable dust mass is
$M_{ \rm d}(t)/M_{\rm ini} = \mu_D/\beta$ and for
$\beta t \ll \beta t_0 \equiv M_{\rm ini}/\psi_{\rm ini}$, 
$M_{ \rm d}(t) = \mu_D \psi_{\rm ini} t$
\citep[see also][]{2011A&ARv..19...43G},
i.e., $M_{ \rm d} (t_0)= 0.63 (\mu_D/\beta) M_{\rm ini}$.
For a bottom-light IMF ($\mu_D=0.018$), SNe will turn of order 1\% of the 
initial total mass into dust in the absence of dust destruction ($\beta=1$),
independent of the initial SFR.

During this early phase, the dust mass is proportional to the SFR and the 
duration of the starburst. However, this is not 
the \mdust--SFR relation we are seeking. To reach the dust levels observed, 
the galaxies would have been forming dust at a steady level for 
$t=M_{ \rm d}(t) / (\mu_D \psi_{\rm ini}) \approx$ 4 Gyr (for a Salpeter IMF). 
While in some cases such an interpretation may be viable (e.g., for the
Milky Way, see Hjorth et al.\ 2014), in general this time scale is 
uncomfortably long and the scenario is not expected to give rise to such a 
well-defined relation.

\subsubsection{Late times: quiescent star formation and the slope of the
\mdust-SFR relation}

At late times, star formation will decrease because of the reduced availability 
of gas. As a consequence of the lower SN rate, there is less dust production. 
The dust destruction term is a combination of consumption through astration 
and grain destruction due to SN shocks.

We seek a relation consistent with the \mdust--SFR relation and therefore
make the {\em ansatz}
\begin{equation}
M_d = A \psi^B.
\end{equation}
From this follows that
\begin{equation}
\frac{dM_d}{dt} = -ABk\delta\alpha^{1/k} \psi^{B+1-1/k}.
\end{equation}
Requiring the exponent of $\psi$ in Equations (15) and (21) to be equal, i.e.,
$1=B+1-1/k$, yields $B=1/k$. In other words,
\begin{equation}
M_d = A \psi^{1/k}
\end{equation}
and hence $M_d \propto M_g$, i.e., a constant dust-to-gas ratio, $\eta_d$,
during the late, non-starbursting phase.
Indeed, requiring the prefactors of Equations (15) and (21) to be equal,
$\mu_D - \eta_d(t) \beta = -ABk\delta\alpha^{1/k}$, shows that $\eta_d(t)$ 
must be time independent. The net dust evolution term is negative and equals 
$-A\delta\alpha^{1/k}$.

Note that for $k=1.5$, $M_d \propto \psi^{2/3}$, while for $k=0.9$ one 
retrieves the \citet{2010MNRAS.403.1894D} \mdust--SFR relation, 
$M_d \propto \psi^{1.11}$.

\subsection{Numerical models\label{numerical}}

For illustration we compute full chemical evolution models with an initial 
mass of $M_{\rm ini}= 3\times 10^{11}$ \msun, initial SFRs corresponding to 
a Chabrier SFR of $\psi_{\rm ini} = 1000$ \msunyr, a range of SN dust 
destruction clearing masses ($M_{\rm cl}$=0--1500 \msun), and for Salpeter or 
bottom-light IMFs \citep[for details see][]{2011A&A...528A..13G}.

Regarding the Schmidt--Kennicutt slope (exponent denoted by $N$ for surface 
densities, $\Sigma_{\rm SFR} \propto \Sigma_{\rm mol}^N$), 
\cite{1998ApJ...498..541K} found a slope of $1.4\pm 0.15$ 
\citep[for a recent review, see][]{2012ARA&A..50..531K}. 
\citet{2012ApJ...745...69K} argue on theoretical grounds that the local
volumetric relation between SFR and gas should be 
$\rho_{\rm SFR} \propto \rho_{\rm gas}^{1.5}$. The models of 
\citet{2012ApJ...760L..16R} however suggest the slope is closer to unity at 
high densities or SFRs, due to stellar feedback.
Indeed, \citet{2013A&A...553A.130F} find $N\approx 1$ in resolved galaxies at 
$z\approx 1.2$ while \citet{2013AJ....146...19L} find $N=1\pm 0.15$ for the
relation between molecular mass and SFR in nearby galaxies.
\citet{2014MNRAS.437L..61S} find a non-universal slope whose average is
$N=0.76\pm 0.16 (2\sigma$).  \citet{2008A&A...479..669C} use $k=1$ in their 
models of proto E galaxies. We here compute models with $k$ ranging from 
0.9 to 1.2 and evolve them over 10 Gyr.

We plot the models in Figure~\ref{f:dacunha}. As a benchmark model (thick 
solid curve) we choose a bottom-light IMF, no SN dust destruction, and $k=0.9$ 
to match the observed \mdust--SFR relation. The dashed curve shows the 
corresponding Salpeter IMF model, while the dotted curves show the effect of 
varying the SN dust destruction clearing mass, $M_{\rm cl}$, for a bottom-light
IMF. These models confirm that the power-law slope of the \mdust--SFR relation 
at late times is consistent with $1/k$. The differing IMFs or clearing masses 
primarily affect the normalization of the curves. The IMF has a strong effect 
on the total amount of dust produced, with bottom-light IMFs leading to 
significantly (up to an order of magnitude) higher dust masses, for a fixed 
(Chabrier IMF equivalent) SFR, confirming previous findings 
\citep{2007ApJ...662..927D,2010ApJ...712..942M,2011A&A...528A..13G,2011A&A...528A..14G,2011A&ARv..19...43G,2011MNRAS.416.1916V,2011ApJ...738...36D}.
More massive systems (not shown) also lead to higher dust masses. Thin solid 
lines show the effect of varying $k$ and confirms the expected $1/k$ behavior.
We verified that incorporation of AGB stars does not change the overall
qualitative behaviour of the models presented here.

\section{Discussion\label{discussion}}

As shown by \cite{2010MNRAS.403.1894D}, using \citet{2008A&A...479..669C} 
chemical evolution models, the \mdust--SFR relation can be interpreted as an 
evolutionary sequence. This is partly confirmed by our analysis. In this 
picture, the initial phase is due to a starburst in which dust is built up at 
essentially constant star formation rate. Due to the maximum attainable dust 
mass derived in Section~\ref{early}, the SMG data points are located, on 
average, below the extrapolation of the linear fit to the SDSS data points 
\citep{2010MNRAS.403.1894D}. We note that to reproduce the highest dust 
masses, very efficient formation of dust from the metals produced by SNe is 
required \citep[see also][]{2010ApJ...712..942M,2011A&A...528A..14G}. 
The \citet{2008A&A...479..669C} models do not reproduce such high dust masses.
The high end of the \mdust--SFR relation is therefore further evidence for 
a surprisingly efficient and rapid dust formation process at work, such as 
that inferred from SN 1987A \citep{2011Sci...333.1258M,2014ApJ...782L...2I}. 
Presumably the dust is either formed directly in SNe or through rapid 
subsequent grain growth in the ejecta/remnant or the ISM, such that the 
majority of the refractory elements available are turned into dust. We note 
that, for the highest dust masses, significant SN dust destruction is not 
allowed by the models.

The numerical models also confirm that the late evolution is characterized by 
a joint decay in dust mass and SFR rate which leads to a power-law relation 
between them, with a slope of $1/k$. Hence, the slope of the \mdust--SFR 
relation can be directly related to the global incarnation 
(Equation~\ref{schmidt}) of the Schmidt--Kennicutt star-formation law 
\citep{1959ApJ...129..243S,1998ApJ...498..541K}. We note that the commonly 
adopted value of $k=1.5$ is not favored by the models which seem to prefer 
values around $1\pm 0.2$.
The joint decay in these quantities is due to a delicate balance between
continued star and dust formation and the parallel consumption of the 
available gas (and dust) reservoir and through dust destruction by SN shocks. 
The balance is characterized by a roughly constant dust-to-gas ratio along the 
\mdust--SFR relation.

We may consider the early and late phases to be two stages of star formation; 
the first being the rapid starburst when all the dust is formed, the second 
being quiescently star-forming galaxies. Of course, the ``initial'' starburst 
in the chemical evolution model need not necessarily refer to the very 
beginning of the evolution of a galaxy -- it may for example relate to the
time when a lot of gas ($M_{\rm ini}$) is supplied to the galaxy (e.g., through 
a merger). It is also possible that some galaxies undergo several starbursts. 
Other processes may be at work in shaping the \mdust--SFR diagram. Quenching of 
star formation will lead to a rapid decline in the SFR. If this is due to 
removal of cold gas and dust from the galaxy \citep[e.g., due to heating or 
expulsion,][]{2008ApJS..175..356H}, a parallel decline in the dust mass will 
lead to a transition more or less parallel to the \mdust--SFR relation. 
However, if the star formation is quenched but the dust retained \citep[e.g., 
when the cold gas reservoir, such as a gaseous galactic disk, becomes stable 
against fragmentation to bound clumps, so-called morphological quenching, 
see][]{2009ApJ...707..250M, 2013arXiv1310.3838G} then a horizontal transition 
is expected. Conversely, merging will produce a parallel upwards evolution 
along the \mdust--SFR relation, perhaps leading to a subsequent starburst.

We stress that the evolutionary models plotted in Figure~1 are for illustration
and are not intended to account for all the data points being part of a single
evolutionary sequence. 
Notably, the 
{\it Herschel} ETGs and passive spirals (red symbols) appear, on average, to 
be offset from the relation (blue points), with higher dust masses at a given 
SFR, or, equivalently, lower SFR at a given dust mass. Part of the effect may 
be due to their average higher masses, a selection effect in that they are 
{\em Herschel} detected, or morphological quenching as discussed above. 
Long-time sustained low star formation from a large gas reservoir appears 
unlikely to account for the full effect given they are early-type galaxies. It 
is interesting to note that the offsets of the {\it Herschel} ETGs and the SMGs 
from the main relation are reminiscent of the similar offsets of such galaxies 
from the main sequence of star formation, i.e., the \mstar--SFR relation.
Possibly, the SMGs and the {\it Herschel} ETGs form a separate evolutionary 
sequence, characterized by a higher $k$.

As suggested by the models, we expect a population of star-bursting galaxies 
significantly below the relation. Because of the logarithmic scale and the short
amount of time spent in this early phase, such galaxies may be rare. Examples
include the low-metallicity low-redshift dwarf galaxy starbursts I~Zw~18
\citep{2013arXiv1310.4842F} and SBS~0035$-$052 \citep{2014A&A...561A..49H}. At
higher redshifts, entering the era of reionization, we expect most galaxies to
have low dust content because of the limited time available since Big Bang and
the onset of star formation. ALMA should uncover a significant population of 
very high redshift galaxies below the local relation. The $z\sim 6.6$ galaxy 
`Himiko' may be one such candidate \citep{2013ApJ...778..102O}.

Given the heterogeneous data sets used, as well as the possible range of model 
parameters
(initial mass, SFR, and IMF in particular) and evolutionary processes entering, 
one expects quite some scatter in the relation, as revealed here. More 
homogeneous samples would be needed to model evolutionary sequences in more 
detail.

\acknowledgments
We thank Haley Gomez, Julie Wardlow, Sune Toft, Stefano Zibetti, Anna Gallazzi,
and Darach Watson
for discussions, and Elisabete da Cunha and Mark Swinbank for making their data 
points available in electronic form.
The anonymous referee provided very insightful comments.
C.G. was supported from the NASA Postdoctoral Program (NPP) and acknowledges 
funding provided by the Danish Agency for Science and Technology and Innovation.
The Dark Cosmology Centre is funded by the Danish National Research Foundation.

\clearpage

\clearpage


\begin{thebibliography}{42}
\expandafter\ifx\csname natexlab\endcsname\relax\def\natexlab#1{#1}\fi

\bibitem[{{Calura} {et~al.}(2008){Calura}, {Pipino}, \&
  {Matteucci}}]{2008A&A...479..669C}
{Calura}, F., {Pipino}, A., \& {Matteucci}, F. 2008, \aap, 479, 669

\bibitem[{{Chabrier}(2003)}]{2003PASP..115..763C}
{Chabrier}, G. 2003, \pasp, 115, 763

\bibitem[{{da Cunha} {et~al.}(2008){da Cunha}, {Charlot}, \&
  {Elbaz}}]{2008MNRAS.388.1595D}
{da Cunha}, E., {Charlot}, S., \& {Elbaz}, D. 2008, \mnras, 388, 1595

\bibitem[{{da Cunha} {et~al.}(2010){da Cunha}, {Eminian}, {Charlot}, \&
  {Blaizot}}]{2010MNRAS.403.1894D}
{da Cunha}, E., {Eminian}, C., {Charlot}, S., \& {Blaizot}, J. 2010, \mnras,
  403, 1894

\bibitem[{{Daddi} {et~al.}(2007){Daddi}, {Dickinson}, {Morrison}, {Chary},
  {Cimatti}, {Elbaz}, {Frayer}, {Renzini}, {Pope}, {Alexander}, {Bauer},
  {Giavalisco}, {Huynh}, {Kurk}, \& {Mignoli}}]{2007ApJ...670..156D}
{Daddi}, E., {et~al.} 2007, \apj, 670, 156

\bibitem[{{Dwek}(1998)}]{1998ApJ...501..643D}
{Dwek}, E. 1998, \apj, 501, 643

\bibitem[{{Dwek} {et~al.}(2007){Dwek}, {Galliano}, \&
  {Jones}}]{2007ApJ...662..927D}
{Dwek}, E., {Galliano}, F., \& {Jones}, A.~P. 2007, \apj, 662, 927

\bibitem[{{Dwek} {et~al.}(2011){Dwek}, {Staguhn}, {Arendt}, {Capak}, {Kovacs},
  {Benford}, {Fixsen}, {Karim}, {Leclercq}, {Maher}, {Moseley}, {Schinnerer},
  \& {Sharp}}]{2011ApJ...738...36D}
{Dwek}, E., {et~al.} 2011, \apj, 738, 36

\bibitem[{{Edmunds}(2001)}]{2001MNRAS.328..223E}
{Edmunds}, M.~G. 2001, \mnras, 328, 223

\bibitem[{{Fisher} {et~al.}(2013){Fisher}, {Bolatto}, {Herrera-Camus},
  {Draine}, {Donaldson}, {Walter}, {Sandstrom}, {Leroy}, {Cannon}, \&
  {Gordon}}]{2013arXiv1310.4842F}
{Fisher}, D.~B., {et~al.} 2013, arXiv:1310.4842

\bibitem[{{Freundlich} {et~al.}(2013){Freundlich}, {Combes}, {Tacconi},
  {Cooper}, {Genzel}, {Neri}, {Bolatto}, {Bournaud}, {Burkert}, {Cox}, {Davis},
  {F{\"o}rster Schreiber}, {Garcia-Burillo}, {Gracia-Carpio}, {Lutz}, {Naab},
  {Newman}, {Sternberg}, \& {Weiner}}]{2013A&A...553A.130F}
{Freundlich}, J., {et~al.} 2013, \aap, 553, A130

\bibitem[{{Gall} {et~al.}(2011{\natexlab{a}}){Gall}, {Andersen}, \&
  {Hjorth}}]{2011A&A...528A..13G}
{Gall}, C., {Andersen}, A.~C., \& {Hjorth}, J. 2011{\natexlab{a}}, \aap, 528,
  A13

\bibitem[{{Gall} {et~al.}(2011{\natexlab{b}}){Gall}, {Andersen}, \&
  {Hjorth}}]{2011A&A...528A..14G}
---. 2011{\natexlab{b}}, \aap, 528, A14

\bibitem[{{Gall} {et~al.}(2011{\natexlab{c}}){Gall}, {Hjorth}, \&
  {Andersen}}]{2011A&ARv..19...43G}
{Gall}, C., {Hjorth}, J., \& {Andersen}, A.~C. 2011{\natexlab{c}}, \aapr, 19,
  43

\bibitem[{{Genzel} {et~al.}(2013){Genzel}, {F{\"o}rster Schreiber}, {Lang},
  {Tacchella}, {Tacconi}, {Wuyts}, {Bandara}, {Burkert}, {Buschkamp},
  {Carollo}, {Cresci}, {Davies}, {Eisenhauer}, {Hicks}, {Kurk}, {Lilly},
  {Lutz}, {Mancini}, {Naab}, {Newman}, {Peng}, {Renzini}, {Shapiro Griffin},
  {Sternberg}, {Vergani}, {Wisnioski}, {Wuyts}, \&
  {Zamorani}}]{2013arXiv1310.3838G}
{Genzel}, R., {et~al.} 2013, arXiv:1310.3838

\bibitem[{{Hjorth} {et~al.}(2014)}]{Hjorth14}
{Hjorth}, J., {et~al.} 2014, {in preparation}

\bibitem[{{Hopkins} {et~al.}(2008){Hopkins}, {Hernquist}, {Cox}, \& {Kere{\v
  s}}}]{2008ApJS..175..356H}
{Hopkins}, P.~F., {Hernquist}, L., {Cox}, T.~J., \& {Kere{\v s}}, D. 2008,
  \apjs, 175, 356

\bibitem[{{Hunt} {et~al.}(2014){Hunt}, {Testi}, {Casasola},
  {Garc{\'{\i}}a-Burillo}, {Combes}, {Nikutta}, {Caselli}, {Henkel},
  {Maiolino}, {Menten}, {Sauvage}, \& {Weiss}}]{2014A&A...561A..49H}
{Hunt}, L.~K., {et~al.} 2014, \aap, 561, A49

\bibitem[{{Indebetouw} {et~al.}(2014){Indebetouw}, {Matsuura}, {Dwek},
  {Zanardo}, {Barlow}, {Baes}, {Bouchet}, {Burrows}, {Chevalier}, {Clayton},
  {Fransson}, {Gaensler}, {Kirshner}, {Laki{\'c}evi{\'c}}, {Long}, {Lundqvist},
  {Mart{\'{\i}}-Vidal}, {Marcaide}, {McCray}, {Meixner}, {Ng}, {Park},
  {Sonneborn}, {Staveley-Smith}, {Vlahakis}, \& {van
  Loon}}]{2014ApJ...782L...2I}
{Indebetouw}, R., {et~al.} 2014, \apjl, 782, L2

\bibitem[{{Jones} {et~al.}(1996){Jones}, {Tielens}, \&
  {Hollenbach}}]{1996ApJ...469..740J}
{Jones}, A.~P., {Tielens}, A.~G.~G.~M., \& {Hollenbach}, D.~J. 1996, \apj, 469,
  740

\bibitem[{{Kennicutt} \& {Evans}(2012)}]{2012ARA&A..50..531K}
{Kennicutt}, R.~C., \& {Evans}, N.~J. 2012, \araa, 50, 531

\bibitem[{{Kennicutt}(1998)}]{1998ApJ...498..541K}
{Kennicutt}, Jr., R.~C. 1998, \apj, 498, 541

\bibitem[{{Krumholz} {et~al.}(2012){Krumholz}, {Dekel}, \&
  {McKee}}]{2012ApJ...745...69K}
{Krumholz}, M.~R., {Dekel}, A., \& {McKee}, C.~F. 2012, \apj, 745, 69

\bibitem[{{Leroy} {et~al.}(2013){Leroy}, {Walter}, {Sandstrom}, {Schruba},
  {Munoz-Mateos}, {Bigiel}, {Bolatto}, {Brinks}, {de Blok}, {Meidt}, {Rix},
  {Rosolowsky}, {Schinnerer}, {Schuster}, \& {Usero}}]{2013AJ....146...19L}
{Leroy}, A.~K., {et~al.} 2013, \aj, 146, 19

\bibitem[{{Martig} {et~al.}(2009){Martig}, {Bournaud}, {Teyssier}, \&
  {Dekel}}]{2009ApJ...707..250M}
{Martig}, M., {Bournaud}, F., {Teyssier}, R., \& {Dekel}, A. 2009, \apj, 707,
  250

\bibitem[{{Matsuura} {et~al.}(2011){Matsuura}, {Dwek}, {Meixner}, {Otsuka},
  {Babler}, {Barlow}, {Roman-Duval}, {Engelbracht}, {Sandstrom},
  {Laki{\'c}evi{\'c}}, {van Loon}, {Sonneborn}, {Clayton}, {Long}, {Lundqvist},
  {Nozawa}, {Gordon}, {Hony}, {Panuzzo}, {Okumura}, {Misselt}, {Montiel}, \&
  {Sauvage}}]{2011Sci...333.1258M}
{Matsuura}, M., {et~al.} 2011, Science, 333, 1258

\bibitem[{{McKee}(1989)}]{1989IAUS..135..431M}
{McKee}, C. 1989, in IAU Symposium, Vol. 135, Interstellar Dust, ed. L.~J.
  {Allamandola} \& A.~G.~G.~M. {Tielens}, 431

\bibitem[{{Micha{\l}owski} {et~al.}(2010{\natexlab{a}}){Micha{\l}owski},
  {Murphy}, {Hjorth}, {Watson}, {Gall}, \& {Dunlop}}]{2010A&A...522A..15M}
{Micha{\l}owski}, M.~J., {Murphy}, E.~J., {Hjorth}, J., {Watson}, D., {Gall},
  C., \& {Dunlop}, J.~S. 2010{\natexlab{a}}, \aap, 522, A15

\bibitem[{{Micha{\l}owski} {et~al.}(2010{\natexlab{b}}){Micha{\l}owski},
  {Watson}, \& {Hjorth}}]{2010ApJ...712..942M}
{Micha{\l}owski}, M.~J., {Watson}, D., \& {Hjorth}, J. 2010{\natexlab{b}},
  \apj, 712, 942

\bibitem[{{Morgan} \& {Edmunds}(2003)}]{2003MNRAS.343..427M}
{Morgan}, H.~L., \& {Edmunds}, M.~G. 2003, \mnras, 343, 427

\bibitem[{{Noeske} {et~al.}(2007){Noeske}, {Weiner}, {Faber}, {Papovich},
  {Koo}, {Somerville}, {Bundy}, {Conselice}, {Newman}, {Schiminovich}, {Le
  Floc'h}, {Coil}, {Rieke}, {Lotz}, {Primack}, {Barmby}, {Cooper}, {Davis},
  {Ellis}, {Fazio}, {Guhathakurta}, {Huang}, {Kassin}, {Martin}, {Phillips},
  {Rich}, {Small}, {Willmer}, \& {Wilson}}]{2007ApJ...660L..43N}
{Noeske}, K.~G., {et~al.} 2007, \apjl, 660, L43

\bibitem[{{Ouchi} {et~al.}(2013){Ouchi}, {Ellis}, {Ono}, {Nakanishi}, {Kohno},
  {Momose}, {Kurono}, {Ashby}, {Shimasaku}, {Willner}, {Fazio}, {Tamura}, \&
  {Iono}}]{2013ApJ...778..102O}
{Ouchi}, M., {et~al.} 2013, \apj, 778, 102

\bibitem[{{Renaud} {et~al.}(2012){Renaud}, {Kraljic}, \&
  {Bournaud}}]{2012ApJ...760L..16R}
{Renaud}, F., {Kraljic}, K., \& {Bournaud}, F. 2012, \apjl, 760, L16

\bibitem[{{Rowlands} {et~al.}(2012){Rowlands}, {Dunne}, {Maddox}, {Bourne},
  {Gomez}, {Kaviraj}, {Bamford}, {Brough}, {Charlot}, {da Cunha}, {Driver},
  {Eales}, {Hopkins}, {Kelvin}, {Nichol}, {Sansom}, {Sharp}, {Smith}, {Temi},
  {van der Werf}, {Baes}, {Cava}, {Cooray}, {Croom}, {Dariush}, {de Zotti},
  {Dye}, {Fritz}, {Hopwood}, {Ibar}, {Ivison}, {Liske}, {Loveday}, {Madore},
  {Norberg}, {Popescu}, {Rigby}, {Robotham}, {Rodighiero}, {Seibert}, \&
  {Tuffs}}]{2012MNRAS.419.2545R}
{Rowlands}, K., {et~al.} 2012, \mnras, 419, 2545

\bibitem[{{Schmidt}(1959)}]{1959ApJ...129..243S}
{Schmidt}, M. 1959, \apj, 129, 243

\bibitem[{{Shetty} {et~al.}(2014){Shetty}, {Kelly}, {Rahman}, {Bigiel},
  {Bolatto}, {Clark}, {Klessen}, \& {Konstandin}}]{2014MNRAS.437L..61S}
{Shetty}, R., {Kelly}, B.~C., {Rahman}, N., {Bigiel}, F., {Bolatto}, A.~D.,
  {Clark}, P.~C., {Klessen}, R.~S., \& {Konstandin}, L.~K. 2014, \mnras, 437,
  L61

\bibitem[{{Smith} {et~al.}(2012){Smith}, {Dunne}, {da Cunha}, {Rowlands},
  {Maddox}, {Gomez}, {Bonfield}, {Charlot}, {Driver}, {Popescu}, {Tuffs},
  {Dunlop}, {Jarvis}, {Seymour}, {Symeonidis}, {Baes}, {Bourne}, {Clements},
  {Cooray}, {De Zotti}, {Dye}, {Eales}, {Scott}, {Verma}, {van der Werf},
  {Andrae}, {Auld}, {Buttiglione}, {Cava}, {Dariush}, {Fritz}, {Hopwood},
  {Ibar}, {Ivison}, {Kelvin}, {Madore}, {Pohlen}, {Rigby}, {Robotham},
  {Seibert}, \& {Temi}}]{2012MNRAS.427..703S}
{Smith}, D.~J.~B., {et~al.} 2012, \mnras, 427, 703

\bibitem[{{Swinbank} {et~al.}(2013){Swinbank}, {Simpson}, {Smail}, {Harrison},
  {Hodge}, {Karim}, {Walter}, {Alexander}, {Brandt}, {de Breuck}, {da Cunha},
  {Chapman}, {Coppin}, {Danielson}, {Dannerbauer}, {Decarli}, {Greve},
  {Ivison}, {Knudsen}, {Lagos}, {Schinnerer}, {Thomson}, {Wardlow}, {Weiss}, \&
  {van der Werf}}]{2013arXiv1310.6362S}
{Swinbank}, M., {et~al.} 2013, arXiv:1310.6362

\bibitem[{{Tielens}(2005)}]{2005pcim.book.....T}
{Tielens}, A.~G.~G.~M. 2005, {The Physics and Chemistry of the Interstellar
  Medium}

\bibitem[{{Valiante} {et~al.}(2011){Valiante}, {Schneider}, {Salvadori}, \&
  {Bianchi}}]{2011MNRAS.416.1916V}
{Valiante}, R., {Schneider}, R., {Salvadori}, S., \& {Bianchi}, S. 2011,
  \mnras, 416, 1916

\bibitem[{{Wuyts} {et~al.}(2011){Wuyts}, {F{\"o}rster Schreiber}, {van der
  Wel}, {Magnelli}, {Guo}, {Genzel}, {Lutz}, {Aussel}, {Barro}, {Berta},
  {Cava}, {Graci{\'a}-Carpio}, {Hathi}, {Huang}, {Kocevski}, {Koekemoer},
  {Lee}, {Le Floc'h}, {McGrath}, {Nordon}, {Popesso}, {Pozzi}, {Riguccini},
  {Rodighiero}, {Saintonge}, \& {Tacconi}}]{2011ApJ...742...96W}
{Wuyts}, S., {et~al.} 2011, \apj, 742, 96

\bibitem[{{Zhukovska} {et~al.}(2008){Zhukovska}, {Gail}, \&
  {Trieloff}}]{2008A&A...479..453Z}
{Zhukovska}, S., {Gail}, H.-P., \& {Trieloff}, M. 2008, \aap, 479, 453

\end{thebibliography}
\end{document}